\documentclass[fleqn,twoside]{article}
\usepackage{espcrc2}

\usepackage{graphicx}
\usepackage[figuresright]{rotating}

\newcommand{\AmS}{{\protect\the\textfont2
  A\kern-.1667em\lower.5ex\hbox{M}\kern-.125emS}}

\title{On the primary particle type of the most energetic Fly's Eye event}

\author{
M.~Risse\address[IFJ]{Institute of Nuclear Physics PAN,
ul.~Radzikowskiego 152,
31-342 Krak\'ow, Poland}\address[FZK]{Forschungszentrum Karlsruhe,
Institut f\"ur Kernphysik, 76021 Karlsruhe, 
Germany}\address{Electronic address: markus.risse@ik.fzk.de},
P.~Homola\addressmark[IFJ],
R.~Engel\addressmark[FZK],
D.~G\'ora\addressmark[IFJ],
D.~Heck\addressmark[FZK],
J.~P\c{e}kala\addressmark[IFJ], 
B.~Wilczy\'nska\addressmark[IFJ]
and 
H.~Wilczy\'nski\addressmark[IFJ]}

\begin{document}

\begin{abstract}
The longitudinal profile of the 320~EeV event observed by the 
Fly's Eye experiment is analysed.
A method of testing the hypothesis of a specific primary particle type
is described. 
Results for different particle types are summarized.
For hadronic primaries between proton and iron nuclei, the
discrepancy between observed and simulated profiles is in the
range of 0.6-1.0$\sigma$ for two different hadronic interaction models
investigated.
For primary photons, the discrepancy is 1.5$\sigma$ assuming 
a standard extrapolation of the photonuclear cross-section with energy.
Larger values of the cross-section at highest energies make primary
photon showers more similar to hadron-initiated events.
The influence of varying the extrapolation of the photonuclear 
cross-section is studied.

\end{abstract}

\maketitle

\section{Introduction}

Deciphering the primary particle type of the most-energetic cosmic rays
would be an important step towards understanding their origin.
Due to the stochastic process of air shower development, it is
in general not possible to unambiguously identify the nature of the 
primary.
However, if the expectation for a specific primary type were found to
be inconsistent with the data, an {\it exclusion} of such a primary
particle hypothesis might be possible.

In this work, we summarize an analysis method developed for this
purpose:
By comparing large statistics simulations to the 
data of an individual event,
the level of consistency between data and the showers simulated for
a given primary particle type is evaluated.
Results of applying this method to the 
320~EeV Fly's Eye event~\cite{flyseye} are presented.
A detailed discussion is given in~\cite{fe04}.

Of particular interest is the primary photon hypothesis.
For the simulation of primary photon showers, the photonuclear 
cross-section $\sigma_{\gamma - air}$ has to be extrapolated to
highest energies.
It is quantified in this work, how the uncertainty in extrapolating 
$\sigma_{\gamma - air}$ influences shower features of primary photons.

\section{Method and application}

Given the reconstructed primary energy, direction, and depth of shower
maximum $X_{max}$ of the 320~EeV Fly's Eye event~\cite{flyseye},
for each assumption on the primary particle type a large set of
detailed Monte Carlo simulations is generated.
The simulations are performed adopting the
primary energy, direction, and local geomagnetic field conditions
as for the observed event.

The simulated $X_{max}$ distribution is then compared to
the measured $X_{max}$ value. The probability of each primary particle
hypothesis to be consistent with the data is
determined, taking the uncertainty in the measured $X_{max}$ into account.
In this way, i.e.~by directly comparing each simulated $X_{max}$
to the data, also non-Gaussian shower fluctuations are preserved
and naturally considered in the probability evaluation.

\begin{figure}[t]
\vspace{-0.5cm}
\begin{center}
\includegraphics[width=0.5\textwidth,angle=0]{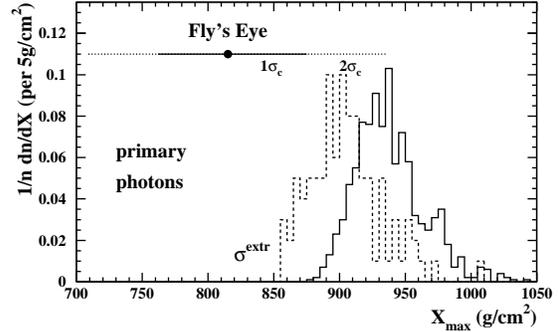}
\end{center}
\vspace{-1.0cm}
\caption{Shower maximum distribution of primary photons
compared to the reconstructed value of the Fly's Eye event.
The measured depth is shown with the $1\sigma$- and
$2\sigma$- uncertainty.
In addition to the simulations with standard photonuclear
cross-section,
results when assuming the extrapolation $\sigma^{extr}$
(see Fig.~\ref{fig-xsec}) are given.}
\label{fig-xmax}
\end{figure}

As an example, in Fig.~\ref{fig-xmax}, $X_{max}$ values for the
primary photon assumption are shown together with the measurement
of $X_{max} = 815\pm 60$~g~cm$^{-2}$.
The simulations are performed with the PRESHOWER code~\cite{preshower}
linked to CORSIKA~6.16~\cite{corsika}.
The discrepancy between primary photons and data obtained this way
from comparing $X_{max}$ is below the 2$\sigma$ level.
The average values $<X_{max}>$ also for hadronic primaries are
listed in Tab.~\ref{tab-xmax}.

\begin{table}[h]
\begin{center}
\caption{Average depth of shower maximum $X_{max}$ and RMS
(both in g~cm$^{-2}$) for the simulated profiles.
 The reconstructed depth of shower maximum
of the 320~EeV Fly's Eye event is 815~g~cm$^{-2}$ with a
combined uncertainty of 60~g~cm$^{-2}$.}
\label{tab-xmax}
\vskip 0.5 cm
\begin{tabular}{lccc}
\hline
&        &  QGSJET~01  &  SIBYLL~2.1 \\
& photon & ~p~~~C~~~Fe & ~p~~~C~~~Fe \\
$X_{max}$ & 937 & 848 808 783 & 882 824 785 \\
RMS       & 26 & 54~ 30~ 22 & 47~  27~  19
\\
\hline
\\
\end{tabular}
\end{center}
\end{table}

\begin{figure}[t]
\vspace{-0.5cm}
\begin{center}
\includegraphics[width=0.5\textwidth,angle=0]{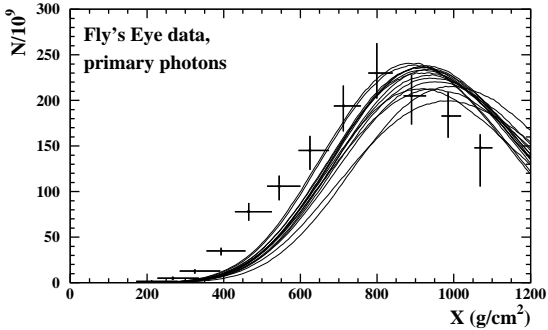}
\end{center}
\vspace{-1.0cm}
\caption{Typical longitudinal profiles of photon-initiated showers
compared to the data.}
\label{fig-profile}
\end{figure}

For the Fly's Eye event, not only $X_{max}$, but the complete
observed profile can be compared to the simulations, see 
Fig.~\ref{fig-profile}.
The statistical method for evaluating the consistency of
simulations and data based on the profile information was
developed in~\cite{fe04}.
An important issue in this comparison is to take the correlation
of the reconstructed profile points in atmospheric depth $X$ into
account, since otherwise a too large discrepancy between
simulation and data might erroneously lead to the conclusion that
the considered primary hypothesis could be rejected.
The results in case of the Fly's Eye event for primary photons and
different hadronic primaries, simulated with the hadronic interaction
models QGSJET~01~\cite{qgsjet} and SIBYLL~2.1~\cite{sibyll},
are lised in Tab.~\ref{tab-ptot}.
The discrepancy of photon shower profiles
 to data is 1.5$\sigma$, and the values
for primaries between proton and iron nuclei range between
0.6-1.0$\sigma$.
Thus, no considered primary particle hypothesis can be rejected for
the Fly's Eye event.

\begin{table}[t]
\begin{center}
\caption{Probability $P$ of a given primary particle hypothesis to be
consistent with the observed Fly's Eye event profile and corresponding
discrepancy $\Delta$ in units of standard deviations.}
\label{tab-ptot}
\vskip 0.5 cm
\begin{tabular}{lccc}
\hline
&        &  QGSJET~01  &  SIBYLL~2.1 \\
         & photon & p~~~~C~~~~Fe & p~~~~C~~~~Fe \\
$P$ [\%] & 13     & ~43~~~54~~~53~ & 31~~~52~~~54
\\
$\Delta$ [$\sigma$] & 1.5 & 0.8~~0.6~~0.6 & 1.0~~0.6~~0.6
\\
\hline
\\
\end{tabular}
\end{center}
\end{table}

\section{Uncertainty of photonuclear cross-section}

\begin{figure}[t]
\begin{center}
\includegraphics[width=0.44\textwidth,angle=0]{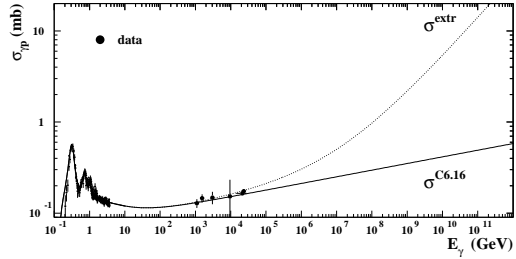}
\end{center}
\vspace{-1.2cm}
\caption{Data~\cite{pdg} and
extrapolations of the photonuclear cross-section
$\sigma_{\gamma-p}$.
Shown are the parametrization used in
CORSIKA~6.16 ($\sigma^{C6.16}$~\cite{stanev})
and a parametrization with an extreme increase of the
cross-sections with energy
($\sigma^{extr}$~\cite{donland}).
The cross-section on air is given by
$\sigma_{\gamma- air}$ = 11.5 $\sigma_{\gamma- p}$.}
\label{fig-xsec}
\end{figure}

The photonuclear cross-section must be extrapolated to highest energies,
i.e.~several orders of magnitude beyond the range of cross-section 
data, see Fig.~\ref{fig-xsec}.
Increased values of $\sigma_{\gamma-air}$ will result in a larger
energy flow from the electromagnetic to the hadronic shower component.
Correspondingly, features of primary photon showers are expected to
change.
More specifically, $<X_{max}>$ becomes smaller and the average
number of muons $<N_{\mu}>$
larger when assuming larger values of $\sigma_{\gamma-air}$ for
primary photon simulations at highest energies.

This has been verified by adopting an extreme parametrization of
the photonuclear cross-section~\cite{donland}, denoted $\sigma^{extr}$
in Fig.~\ref{fig-xsec}.
In that case, $<X_{max}>$ for the Fly's Eye event simulations
is decreased by $\simeq$30~g cm$^{-2}$ (see Fig.~\ref{fig-xmax}).
This corresponds to a smaller discrepancy of the primary photon hypothesis
to data, reduced from 1.5$\sigma$ to about 1.2$\sigma$.
The shift in $<X_{max}>$ is relatively modest here, since for the
conditions of the Fly's Eye event, strong precascading of the initial
photon in the geomagnetic field occurs. The shift is found to
be much larger (up to 100-200~g~cm$^{-2}$) for highest-energy photons
that enter the atmosphere without precascading.

Although the muon content of the shower was not measured by the
Fly's Eye experiment, it is interesting to check for the 
effect in $<N_{\mu}>$.
Applying the extrapolation $\sigma^{extr}$, the muon number on
ground is increased by $\simeq$75\%, thus reducing considerably
the difference in muon content between primary photon and proton
showers.
Therefore, the uncertainty in extrapolating
the photonuclear cross-section to
highest energies must be seriously considered when conclusions on
primary photons are to be drawn.

\section{Conclusion}

Both primary photons and any hadron between proton and iron nuclei
can not be excluded as primary particle of the 320~EeV Fly's Eye event.
The method developed for this analysis is further exploited in an
investigation of the most energetic AGASA events as presented
at this conference~\cite{piotrek}.
The large uncertainty in extrapolating the photonuclear cross-section
can considerably influence the predictions of features of primary
photon events and deserves a further careful study.

{\it Acknowledgments.}
This work was partially supported by the Polish State Committee for
Scientific Research under grants No.~PBZ~KBN~054/P03/2001 and
2P03B~11024 and in Germany by the DAAD under grant No.~PPP~323.
MR is supported by the Alexander von Humboldt-Stiftung.


\begin{thebibliography}{9}


\bibitem{flyseye}
D.J.\,{}Bird\,{}et al., Astrophys.\,{}J.\,{}{\bf 441} (1995) 144

\bibitem{fe04}
M.\,{}Risse\,{}et al., Astrop.\,{}Phys.\,{}{\bf 21} (2004) 479

\bibitem{preshower} P.~Homola et al.,
astro-ph/0311442 (2003)

\bibitem{corsika} D.~Heck et al., Report {\bf FZKA 6019}, For\-schungszentrum
     Karls\-ruhe (1998); D.~Heck and J.~Knapp, Report {\bf FZKA 6097},
		 Forschungszentrum Karls\-ruhe (1998)

\bibitem{qgsjet} N.N.~Kalmykov, S.S.~Ostapchenko, and A.I.~Pavlov, 
		 Nucl.~Phys.~B (Proc.~Suppl.) {\bf 52B} (1997) 17

\bibitem{sibyll}
R.~Engel et al.,
Proc.~26th
Int.~Cosmic Ray Conf., Salt Lake City {\bf 1} (1999) 415


\bibitem{pdg} 
S.\,{}Eidelmann\,{}et al., Phys.\,{}Lett.\,{}B\,{}592\,{}(2004) 1
		    
\bibitem{stanev}
T.~Stanev et al., Phys.~Rev.~D {\bf 32} (1985) 1244

\bibitem{donland}
A.~Donnachie, P.~Landshoff, Phys.~Lett.~B {\bf 518} (2001) 63

\bibitem{piotrek}
P.~Homola et al., to be published in Nucl.~Phys.~B (Proc.~Suppl.)


\end{thebibliography}
\end{document}